Ihor Pysmennyi,
Roman Kyslyi,
Kyrylo Kleshch

# AI-DRIVEN TOOLS IN MODERN SOFTWARE QUALITY ASSURANCE: AN ASSESSMENT OF BENEFITS, CHALLENGES, AND FUTURE DIRECTIONS

*Traditional quality assurance (QA) methods face significant challenges in addressing the complexity, scale, and rapid iteration cycles of modern software systems and are strained by limited resources available, leading to substantial costs associated with poor quality.*

*The object of this research is the quality assurance processes for modern distributed software applications. The subject of the research is the assessment of the benefits, challenges, and prospects of integrating modern AI-oriented tools into quality assurance processes. Comprehensive analysis of implications was performed on both verification and validation processes covering exploratory test analyses, equivalence partitioning and boundary analyses, metamorphic testing, finding inconsistencies in acceptance criteria (AC), static analyses, test case generation, unit test generation, test suit optimization and assessment, end to end scenario execution. End to end regression of sample enterprise application utilizing AI-agents over generated test scenarios was implemented as a proof of concept highlighting practical use of the study. The results, with only 8.3% flaky executions of generated test cases, indicate significant potential for the proposed approaches. However, the study also identified substantial challenges for practical adoption concerning generation of semantically identical coverage, "black box" nature and lack of explainability from state-of-the-art Large Language Models (LLMs), the tendency to correct mutated test cases to match expected results, underscoring the necessity for thorough verification of both generated artifacts and test execution results.*

*The research demonstrates AI's transformative potential for QA but highlights the importance of a strategic approach to implementing these technologies, considering the identified limitations and the need for developing appropriate verification methodologies.*

## 1. Introduction

In today's fast-paced digital landscape, quality assurance (QA) is a critical determinant of a company's reputation, dependability, and overall success. As shown in Fig. 1, recent reports indicate that insufficient software quality cost the US market approximately 2.41 trillion USD in 2022 – a 15% increase since 2020 [1].

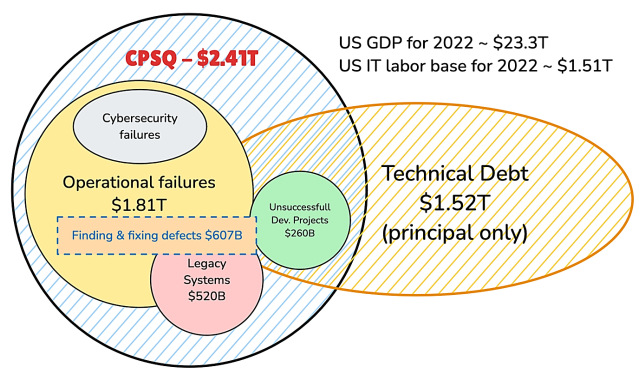

**Fig. 1.** Cost of software quality 2022 [1]

Traditional QA contains chain of multiple labor-intensive practices, which are a common bottleneck for product release lifecycle and struggle to address complexity and scale of modern enterprise software systems, and SAAS platforms. Generating test cases requires comprehensive system analysis for exploratory testing and equivalence partitioning and is frequently shortened to only positive scenarios. Verification of existing functionality is performed via insufficient unit and integration test coverage as well as manual regression testing coupled with often non-reliable, "flaky" end-to-end automation [2]. Defect prioritization techniques are based on business value of affected flows which lead to having huge backlogs of unresolved issues. Mentioned challenges have exposed gaps in current testing approaches, which are often compounded by chronic underinvestment in QA [3].

These complexities, combined with increasingly shorter SDLC iterations, emphasize the importance of "shift left testing" – performing QA activities earlier in the development cycle. Project Management Body of Knowledge (PMBOK) recognizes proactive inspection and review work focused on finding quality issues early in SDLC as good investments [4]. Clear disadvantage of this approach is increased pressure on testing teams caused by short release cycles [5].

Recent breakthroughs in Artificial Intelligence, particularly the advanced capabilities demonstrated by LLMs and AI agentic systems, present a timely opportunity to address these persistent QA challenges. Their potential for automating complex tasks, understanding natural language requirements, generating code, and adapting to dynamic environments offers new possibilities to enhance efficiency and improve confidence in quality where traditional methods fall short. The substantial economic impact of poor software quality further underscores the urgency and relevance of exploring these AI-driven solutions.

Therefore, *the aim of this research* is to provide an assessment of adoption, practical implications, and future directions for leveraging AI in QA processes. The application of AI-driven tools was analyzed across key QA activities, specifically focusing on verification tasks such as:

1. Static analyses.
2. Test case generation.
3. Unit test generation.
4. Test suite optimization and assessment.
5. Test data generation.
6. Regression automation (including agentic regression).

As well as validation tasks including:

1. Exploratory test analyses.
2. Equivalence partitioning.
3. Finding inconsistencies in AC.

Subsequently, it is possible to evaluate the potential benefits and inherent challenges associated with these applications and demonstrated practical feasibility of utilizing AI agents for end-to-end regression over generated test scenarios.

## 2. Materials and Methods

### 2.1. The object of research

*The object of this research* is the quality assurance processes for modern distributed software applications. Tests and QA processes can be classified by implementation complexity and accuracy [6]. This classification is shown in Fig. 2. For example, exhaustive testing provides absolute accuracy as all possible app states are verified but is unfeasible to implement in enterprise applications due to amount of these states a thus having effectively infinite complexity. Strong type checking, such as implemented in C, C++ and Java does not require additional effort from software engineer and can be found on other edge complexity scale but covers much smaller set of potential defects. On accuracy testing techniques can be pessimistic, reporting errors when there are none, such as simplistic program analyses or optimistic – showing no defect when errors are present. For instance, end-to-end automated regression testing by default covers only positive scenarios.

Further in this chapter, let's provide evaluation of implications of AI-driven tools on testing methods above followed by experiments over sample e-commerce application [7] in chapter 3.

### 2.2. Static code analyses

Static code analysis is an essential part of modern dev-ops practices as it allows to proactively detect and prevent errors on the development stage, saving resources on testing. Current tools, such as SonarQube have two limitations though [6]:

1. Unsound analyses is overly optimistic and skips potential defects having negative impact on recall metric.
2. Sound static analyses is pessimistic producing too many false positives which causes distrust to such reports from software engineers (precision metric).

Therefore, to evaluate performance of static analyses tool it is suggested to utilize F-metric.

To increase precision of defect detection, utilization of modern transformer models is proposed [8]. Pre-trained transformers, such as CodeBERT allow achieving 92% F1 scores for NameErrors; 79% for TypeError detection; 66% for IndexError; 66% for IndexError; 95% for AttributeError; 60% for ValueError; 71% for EOFError; 100% for SyntaxError detection and 100% for ModuleNotFoundError. These numbers show significant improvement over baseline F1 score of not pre-configured SonarQube [9].

Another application of AI-enhanced static code analysis is vulnerability detection. It is argued that AI analyses over code structure and semantics results in prediction of probability of various vulnerability types and thus security-related defects [10].

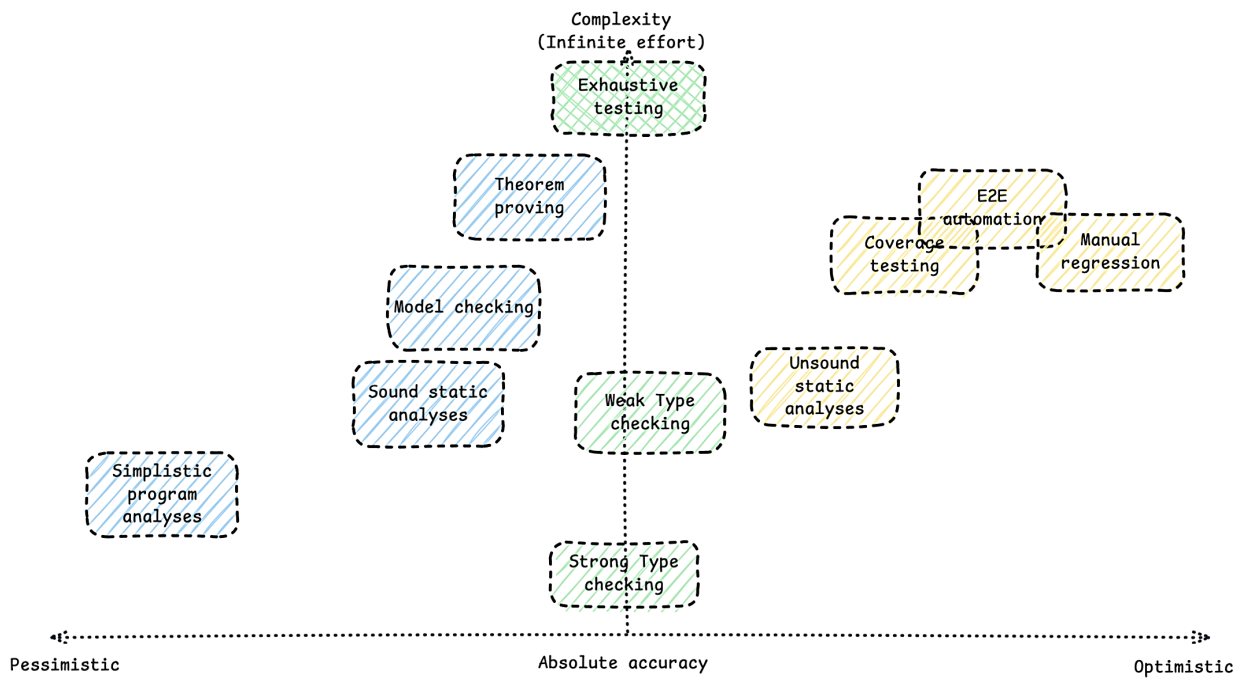

**Fig. 2.** Test analyses square

### 2.3. Test case generation and validation

The foundational part of quality assurance is test case design. A test case is a document that outlines specific conditions and actions to be executed on a software application, aimed at verifying that a particular feature functions as expected and typically consists of [11]:

1. Test case ID and title.
2. Test data – data required to execute a test case. For example, login and password or customer age to fill in UI form to trigger some specific expected validation message.
3. Preconditions – initial state of the system required to execute test case.
4. Steps – an ordered action sequence the tester must perform to execute a test scenario. This executable sequence is called test scenario.
5. Expected results – expected system state and behavior after performing test steps.
6. Postconditions – optional set of steps required to revert system to blank state.

Utilization of large language models (LLMs) contributes to test design in various way such as generation of test cases from existing acceptance criteria or bug reports, generation of documentation, exploratory analyses [12]. Further in experiment section it is possible to prove these scenarios and evaluate LLM-generated-cases with regard to their executability and validity. Executability is a binary metric which describes whether test scenario can be executed without manual changes. Validity – depicting if executing test scenario results in expected behavior in case of working system. According to recent research, LLM-generated test case executability varies from 70% and up to 90% depending on product complexity [13].

Moreover, LLMs can be utilized to evaluate both generated and existing test cases via "LLM-as-a-judge" methodology [14]. As acceptance criteria of user story is frequently not self-sufficient and relies on context or technical documentation it is suggested to use Retrieval-Augmented Generation (RAG) for enriching generation context and having fuller representation of desired functionality.

With integration of enterprise knowledge databases such as Confluence, LLM can perform exploratory analyses and equivalence partitioning. Principal scheme of such AI-assisted test-case generation and validation system is depicted in Fig. 3: user story description and acceptance criteria along with additional retrieved context such as relevant wiki pages on confluence, emails or slack messages are passed into "Test Generator" LLM.

Test generator produces structured output with suggested test-cases which in turn are evaluated with "LLM-as-a-judge" agent. This agent also may use context from external documentation sources as well as verify executability of test case via utilization of state-of-the-art browser agents such as BrowserUse [15] or other similar tools. In case of validation failure report on test-case errors and inconsistencies is backpropagated as additional context to generator agent enabling autocorrection functionality.

### 2.4. Structural test code generation

Structural testing consisting of unit and integration tests is an integral part of SDLC as the only way to prove that software works is verification of particular scenarios, effectively defining coverage as an ultimate quality assurance metric [16]. Despite its paramount importance, 55.6% of developers find their test coverage insufficient, while on average software engineer spends 15.8% of worktime on test coverage [17]. Maintenance complexity, fragility, lack of time, overhead in execution time, difficulty of understanding what needs to be covered are among the top impediments of structural test design and implementation. Another challenge is posed by fact that test results can be non-deterministic: degree of test-outcome variation is referred to as "flakiness" [2].

Further let's examine how utilization of various AI based tools helps mitigating challenges above.

Typical test method consists of two parts: test prefix and test oracle. Test prefix sets up test environment, data and define mocks as and test oracle verifies if behavior of test object under execution equals to expected one, e. g. asserts output and verifies execution flow.

Since recent LLMs have shown remarkable code generation capabilities and even gained wide industry adoption, researchers started extending LLM applications explicitly to unit test generation tasks. Despite initial promising outcomes, generating effective unit tests introduces unique complexities that differ significantly from standard code generation scenarios [18] – such as lack of contextual understanding; redundant cases when generated test scenarios are not extending coverage and partial mocking.

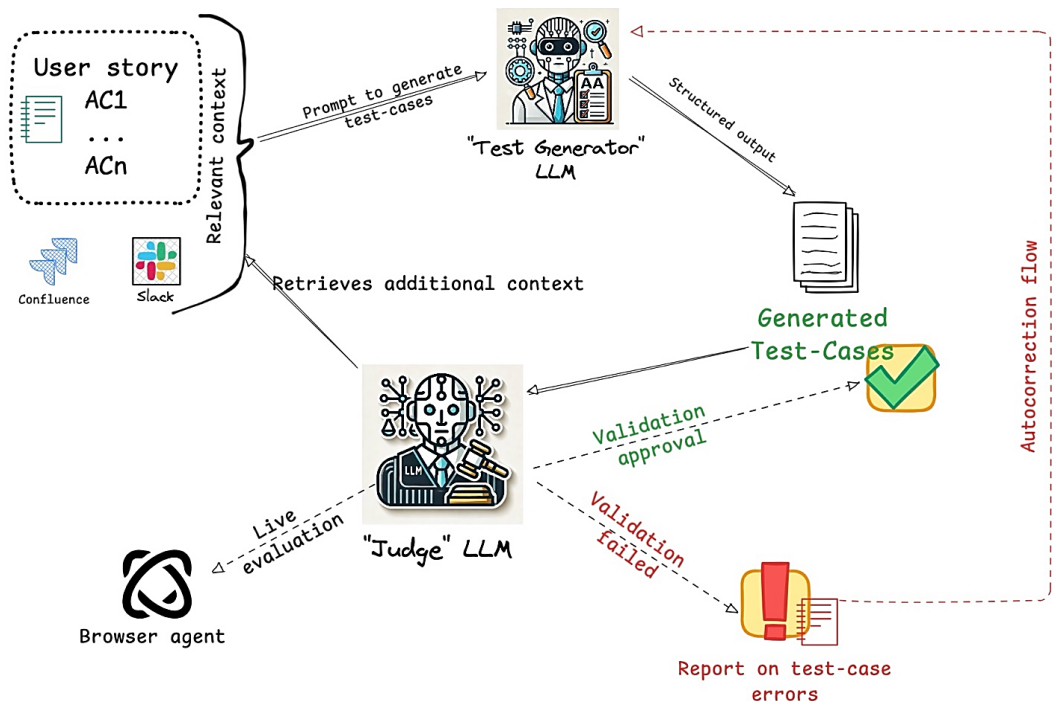

**Fig. 3.** Test-case generation and validation system

To mitigate some of mentioned limitations, Meta inc. offered multi-step observation-based test generation process [19]. Their TestGen tool instruments application at runtime recording real-life parameters and return values of executed functions. The observations, along with the initial application source code, are fed into the "Test Generator" plugin, which constructs the tests. Subsequently, the "Test Publisher" generates the build system files and dependencies for the test cases, executing them five times to ensure they are not flaky. For all tests that pass consistently and demonstrate stability, the "Test Publisher" produces a source code package ready to be included in continuous integration process. Test Generator can have multiple implementations allowing to adjust test generation strategy [19]. Additional advantage of basing test cases on execution observations is ability to create test when expected result is not clear, so called "no oracle" scenario.

As LLMs have probabilistic responses, they may produce the similar output for same prompt over different samples, thus it's important to filter out semantically identical coverage [20]. This can be done either via semantical comparison with existing tests, where embedding-based search plays important role, or by comparing various coverage metrics, mainly statement and branch coverage, before and after inclusion of generated test to test suit.

Further assessment of structural test coverage quality is performed with mutation testing. Small changes called "mutations" are introduced into source code and then ability of test suit to identify those issues is evaluated. Despite its recognized effectiveness in identifying inadequacies in test coverage, mutation testing historically suffers from computational overhead and scalability challenges due to a large number of possible mutants. Utilization of LLM allow focusing mutation on specific aspects of test quality (e. g. security); generate relevant and highly realistic mutations; discover and remove equivalent mutants, and generate unit tests targeted to eliminate discovered mutants [21].

Another promising application of AI in test implementation is enrichment of test generation context. Utilization of RAG with external sources such as GitHub issues featuring detailed code examples, discussions and resolutions significantly increase pass rate of generated unit tests [22].

To address difficulties in setting up test environment proposed dual-environment architecture by introducing LLM-based agent Repo2Run, which allows to fully automate environment configuration and generate executable Dockerfiles for arbitrary Python repositories with 86% success rate [23].

Proposed system for generation and validation of test cases is shown in Fig. 4.

Code generation agent generates tests based on enriched context: original user story or bug report, application source code, execution instrumentation, similar GitHub tickets, etc. Then, generated tests are evaluated regarding their executability, coverage effect and relevancy.

**2.5. End-to-end automation**

From 40 to 85% of faults found in cloud systems are caused by errors in error handling, optimization, configuration, race, hang, space, load which span across different use-cases in contrast to easily detectable with structural tests errors in logic [24]. To detect these bugs End-to-End testing is utilized. This approach covers functional and data flows across multiple subsystems interfacing with each other from start to end usually via interaction with system UI as close as possible to real-world user behavior. Naturally, this approach gives maximum confidence in correctness of test object. However, automation of these test present challenges among which most notable are [25]:

1. Test flakiness caused by timing issues, unexpected popups and browser quirks. This led to big amount of false positive results.

2. Maintenance cost. Modern single page front-end frameworks often generate code impossible to cover with reliable locators and x-paths of page elements change in case of UI framework update which in turn requires significant redesign of end-to-end test code.

3. Dependence on third party for support business flows such as login or payment capabilities. As third-party release cycle is decoupled from client's release cycle, updating third party UI leads to unexpected.

4. Relying on end-to-end regression is wearing out of tests.

As demonstrated in the ReAct framework, empowering LLM with tools to interact with external environment significantly enhance complex reasoning capabilities which is key to simulation of human user behavior and executing end-to-end scenarios [26]. The ReAct approach enables AI agents to reason dynamically about the current UI state and environment conditions and execute automated browser interactions. Thanks to feedback loop from execution environment, this approach is reducing test flakiness caused by timing issues and unexpected browser behaviors. Furthermore, AI agents can dynamically adapt to changes in UI frameworks. They can contextually identify elements, mitigating the maintenance costs typically associated with rigid locators and x-paths, and autonomously manage third-party UI variations. Additionally, LLMs give ability to partially mitigate this factor by generating coherent, yet variable test data for each execution.

AI-agent based end-to-end automation testing system workflow is shown in Fig. 5 and is evaluated in following chapter.

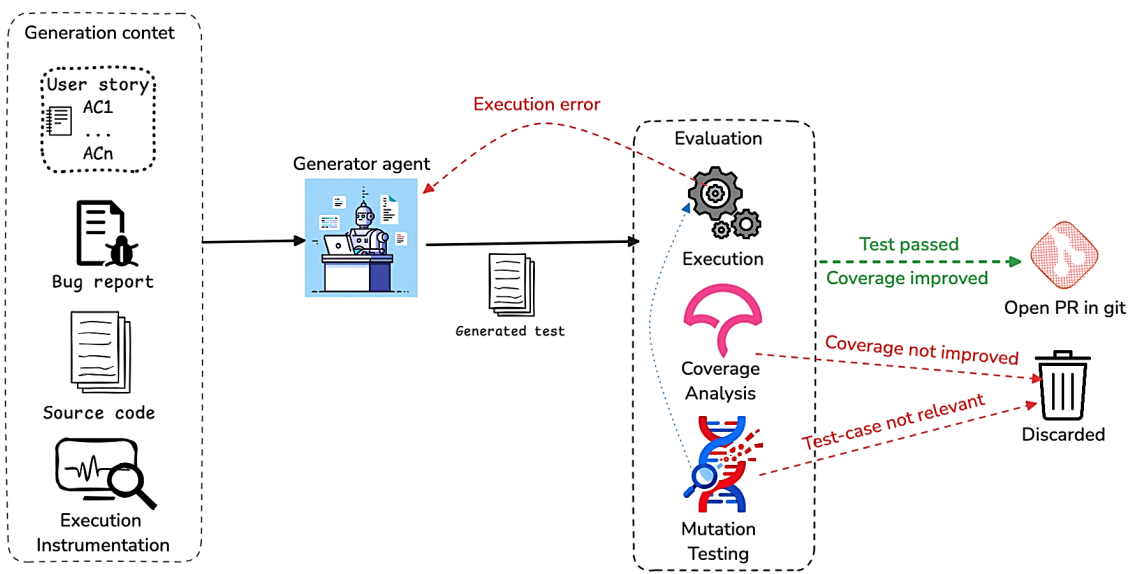

**Fig. 4.** Unit-test generation system

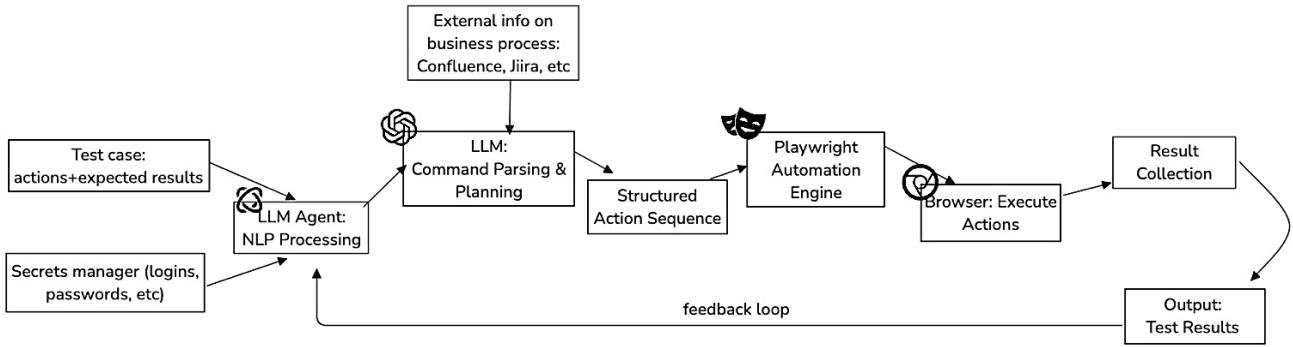

Fig. 5. AI-agent based end-to-end automation testing

Test execution agent receives test case and applicable test data. LLM then reasons over it and test plan with structured action sequence is generated. Agent executes actions and receives actual state from browser via playwright automation engine. This "ground truth" state is backpropagated back to reasoning model and in case of its divergence from expectation corrective set of actions is planned and performed.

### 2.6. Testing AI applications

AI and deep learning in particular have become an integral part of multiple critical software systems, however verification and validation of their behavior pose significant challenge. A key difficulty lies in handling the probabilistic outputs produced by AI models, as slight changes in input data can lead to significant variations in results, complicating the establishment of reliable test oracles. In addition to that single input can have multiple valid outputs making it difficult for tester to define expected behavior. Second issue is lack of explainability, especially in state-of-the-art transformer models which not only hinders debugging efforts but also complicates test case design. Additionally, the evolving nature of AI models complicates maintenance of existing coverage.

One of promising ways to tackle these challenges is metamorphic testing [27, 28]. Let for defined input $x$, AI model generate output $f(x)$ then in case of input transformation $x'$ transformed output $f(x')$ must satisfy expected relation $R$

$$R(f(x), f(x')) = true. \qquad (1)$$

There are three most common metamorphic relations. Invariance is used when transforming input should not affect output. Increase when transforming input should increase output, for instance new record in credit history increase personal credit score. Decrease: for example, adding bankruptcy record should negatively affect credit rating.

Utilization LLMs allow runtime reasoning over expected output transformation over input change. Typical examples of input transformations for various use-cases are shown in Table 1.

Table 1

Metamorphic input transformations

| Input scenario | Input transformations (data augmentation) |
|---|---|
| Computer vision | Blurring, rotation, scaling, light changes, adding noise, image generation |
| Text | Typos, switching sentiment words, using synonyms, adding unrelated text |
| Categorical variable | Switching categories [29] |
| Numeric | Increasing or decreasing |

In Fig. 6 it is possible to show a bug found on EURopean Employment Services job search portal (EURES) with help of metamorphic testing.

With free text search input transformation "*adding restrictive keyword*" expected metamorphic relation for quantity of results is "Decrease". As output number have increased clear fault is observed.

Fig. 6. Metamorphic testing example

## 3. Results and Discussion

### 3.1. Experiment 1. Test case generation

This experiment is dedicated to evaluation of test case executability. Test case generation is performed for sample e-commerce application [7]. Following prompt was used to generate test cases:

*You work as a QA engineer. Generate concise test case in the following format:*
*1. Test case ID and title.*
*2. Test data – data required to execute a test case. For example, login and password or customer age to fill in UI form to trigger some specific expected validation message.*
*3. Preconditions – initial state of the system required to execute test case.*
*4. Steps – an ordered action sequence the tester must perform to execute a test scenario. This executable sequence is called test scenario.*
*5. Expected results – expected system state and behavior after performing test steps.*
*6. Postconditions – optional set of steps required to revert system to blank state.*
*!!!Important test steps should cover only focal scenario and not preconditions*
*For following User-Story:*
*Title: As a user I want to see preview picture on product list to be able to select product.*
*Preconditions: User logged in with username and password.*
*Acceptance criteria: Each product in product list has a distinctive product picture.*

Test cases were generated for 5 user-stories covering login, product list, sorting, checkout and product preview functionalities. Full description of user stories, corresponding acceptance criteria and generated cases can be found in electronic supplementary material [30]. Test cases were evaluated with following metrics:

1. *Executability* – percentage of cases that can be executed without modification.
2. *Coverage* – if generated test cases cover all acceptance criteria specified in user story.
3. *Tokens used* – metric allowing to estimate price of test case generation.

Following state-of-the-art models were evaluated as well as their cost-efficient counterparts: OpenAI gpt-4.5-preview-2025-02-27, gpt-4o-2024-08-06, gpt-4o-mini-2024-07-18; Google gemini-2.5-pro-exp-03-25, gemini-2.0-flash-lite; Mistral: mistral-large-latest (24.11), ministral-8b-latest (24.10). A total of 81 executable test cases were generated for 14 distinct acceptance criteria. Aggregated evaluation outcomes are depicted in Table 2.

As can be seen from results overall LLMs show promising and, in particular cases, production-ready results in generation of test cases. Following challenges were uncovered:

1. Small models struggle to return responses in consistent format. Fine-tuning prompt, such as providing more examples for few-shot learning, help mitigating this issue.

2. GPT and Mistral models frequently confuse test cases and acceptance criteria. While still covering execution steps, they do it in very inconsistent form.

3. There are occurrences of semantically overlapping test-cases across dataset. LLM-as-a-judge can be used to identify and correct these test design errors.

Google gemini-2.5-pro-exp-03-25 demonstrated the best overall results. This model provided comprehensive analysis and reasoning, leading to improvements in the original Acceptance Criteria (AC) and maximization of coverage quality. Notably, the estimated cost for generating the entire test suite remained low at 0.026 USD, or 0.005 USD per user story. In addition to metrics above test cases of this model were of highest quality with respect to independence, clarity and interpretability. Smaller and cost-efficient models ministral-8b and gpt-4o-mini also demonstrated consistent coverage, unlike their alternative from Google gemini-2.0-flash-lite which was failing to produce meaningful outputs.

### 3.2. Experiment 2. Utilization of LLM agents for end-to-end automation

In this experiment, replacing manual or programmed end-to-end regression with browser-enabled AI agent described in section 2.4 was evaluated. It is possible to compare state of the art large and "small" LLMs from OpenAI: their reasoning capabilities in positive and negative scenarios; verify coverage quality via introduction of "mutant test-cases" where a terminal error is introduced, and execution is expected to fail. Previously generated test cases were utilized for building following flows:

1. Login successful flow.
2. Login with non-existing user (negative scenario).
3. Mutation over login successful flow with corrupted data to verify false negative scenario.
4. Complete end-to-end flow of main business process: login, adding goods to cart and checkout processes.
5. Sorting products by price to verify reasoning capabilities of LLM.
6. Mutation over sorting products by price to verify false negative scenario.

Execution of each scenario was performed 4 times per model with total of 48 executions being evaluated.

Below is the prompt used for login flow. Other prompts and execution logs are available as supplementary material [30].

*### Prompt for QA regression agent*

**Table 2**

*\*\*Objective:\*\**

*As a QA engineer performing regression you will make sure that all test flows listed were executed successfully.*
*If test passes you will log Flow name: passed.*
*If test fails or you are not sure if it passes u will log Flow name: failed and step on which it failed. Add possible root cause if you can identify it.*

*\*\*Important:\*\**
*– Mark test as passed only if u a sure that all steps were executed successfully and in accordance with AC*
*– Log all the selectors used throughout each test in format: step_name: action_performed: selector_used*

Evaluation of LLMs for test case generation

| Model name | AC covered | Exploratory testing over stories | Coverage according to format | Executable TC generated | Tokens used |
|---|---|---|---|---|---|
| gemini-2.0-flash-lite | 57% | 20% | 20% | 8 | 9288 |
| gemini-2.5-pro-exp-03-25 | 100% | 60% | 100% | 17 | 7149 |
| ministral-8b-latest (24.10) | 100% | 0% | 20% | 13 | 4631 |
| mistral-large-latest (24.11) | 79% | 0% | 0% | 9 | 5321 |
| gpt-4o-mini-2024-07-18 | 100% | 0% | 20% | 10 | 3582 |
| gpt-4o-2024-08-06 | 86% | 20% | 80% | 11 | 3397 |
| gpt-4.5-preview-2025-02-27 | 100% | 60% | 80% | 13 | 3876 |

---
*** Flow: Order_and_Checkout ***
---
**Steps:**

1. Navigate to the [Sauce Labs Demo website](https://www.saucedemo.com/).
2. Enter 'standard_user' into the Username field.
3. Enter 'secret_sauce' into the Password field.
4. Make sure Username and Password are filled in. If not, fill them in.
5. Click the 'Login' button.

**Expected Result:**
*The user is redirected to the product list page (e. g., `/inventory.html`).
*The product list/inventory container is visible.
**Important:** Use only test data provided in test case. do not use any other extra data.
**Important:** Ensure efficiency and accuracy throughout the process.

As shown in Table 3, both models have shown surprisingly good results in all tested scenarios: only 8.3% of all executions exhibited flaky results which is within the range of current industry statistics [31]. Errors exhibited by GPT-4o-mini did not have any pattern, while GPT-4o consistently failed to reason price comparison in product list.

Worth highlighting that more advanced planning capacity of GPT-4o model resulted in more efficient execution of complex flows covering multiple business processes avoiding locator errors from the first shot (Fig. 7).

Table 3

Performance of LLMs in end-to-end test executions

| Model | Flow name | Average execution time, sec | Average tokens per execution | Average price per execution, USD | Flaky failed executions |
|---|---|---|---|---|---|
| gpt-4o-mini-2024-07-18 | Login | 26 | 113k | 0.02 | 0% |
| | Login negative | 35 | 170k | 0.03 | 0% |
| | Login mutated | 30 | 170k | 0.03 | 0% |
| | Sorting | 40 | 174k | 0.03 | 0% |
| | Sorting mutated | 53 | 232k | 0.03 | 25% |
| | Buys&Checkout | 70 | 350k | 0.05 | 25% |
| gpt-4o-2024-08-06 | Login | 33 | 24k | 0.06 | 0% |
| | Login negative | 33 | 14k | 0.04 | 0% |
| | Login mutated | 30 | 18k | 0.05 | 0% |
| | Sorting | 46 | 25k | 0.06 | 0% |
| | Sorting mutated | 54 | 29k | 0.07 | 50% |
| | Buys&Checkout | 93 | 70k | 0.18 | 0% |

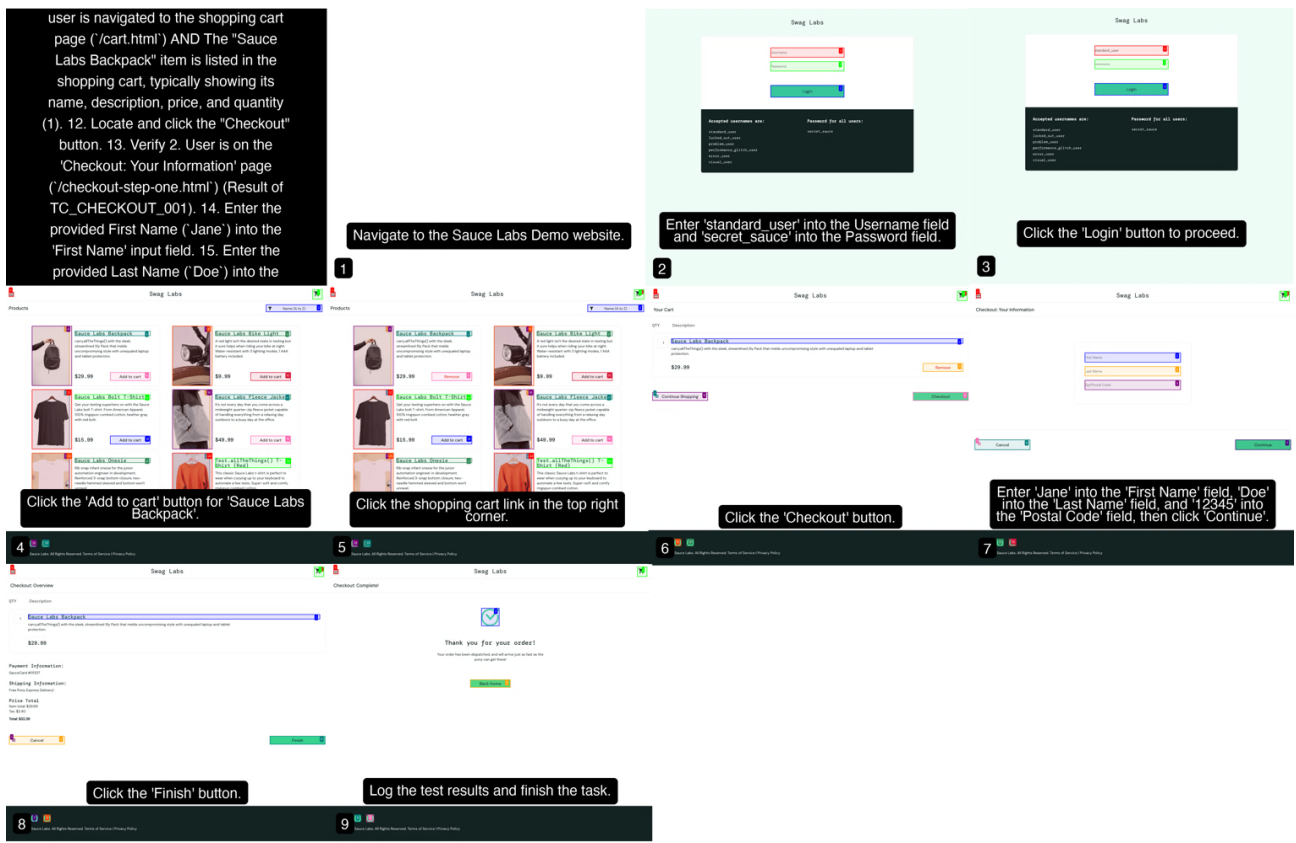

Fig. 7. GPT-4o end-to-end flow

However, as shown in Fig. 8, GPT-4o-mini was able to correct itself after performing observation of state after action according to ReAct framework [26] demonstrating same overall performance with 13.7% benefit in execution time and 142% in cost.

During the execution of the complex end-to-end "Buys&Checkout" scenario, the GPT-4o-mini agent have entered infinite reasoning loop. As depicted in Fig. 9, when attempting to locate and interact with the shopping cart element, the agent entered a repetitive cycle:

1. It attempted an action.
2. Observed an incorrect UI state (e. g., a menu opening instead of navigating to the cart).
3. Inferred failure.
4. Attempted corrective actions (like going back or trying to close the interfering element).
5. Retried the original goal.

While the ReAct framework's observation-action-thought cycle is designed for dynamic adaptation, in this instance with the less capable model it led to agent completely stuck on the same sub-task or for several iterations (Fig. 8). At the same time, the larger GPT-4o model demonstrated smooth execution. This type of iterative struggling highlights potential limitations in the planning or state evaluation capabilities of smaller models. It is aligning with broader observations where LLMs can enter unproductive reasoning cycles or loops when facing complex multi-step tasks, failing to converge on a correct plan or action sequence efficiently [32]. To mitigate potential cost issues due to such errors implementation of model guardrails is advised.

Interesting behavior was observed with regard to mutated test-cases. GPT-4o model was consistently trying to correct mutated test case to achieve expected state. This raises a question of verification of execution logs to avoid false negative results and poses a significant challenge in adoption of AI-agent best regression as part of continuous integration pipeline.

In summary, the experiments presented in this section highlight the tangible practical significance of integrating modern AI tools into QA workflows. The demonstrated ability of leading LLMs to generate comprehensive and largely executable test cases from user stories offers a direct pathway to accelerate test design and potentially improve requirement coverage early in the SDLC. Similarly, the success of AI agents executing complex end-to-end scenarios provides a practical alternative or supplement to traditional automation, offering enhanced resilience to UI changes and the potential to reduce maintenance overhead.

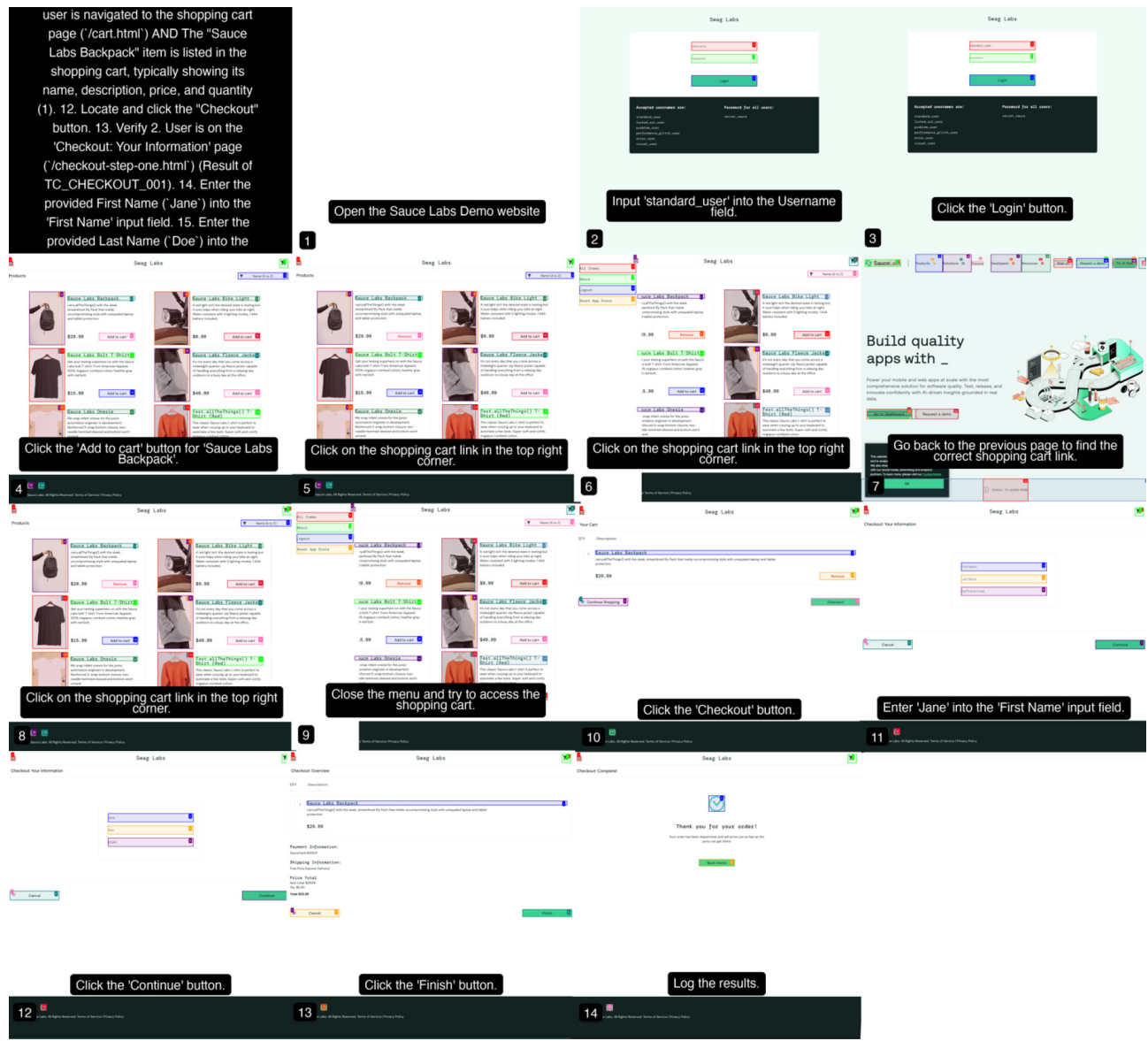

**Fig. 8.** GPT-4o-mini end-to-end flow

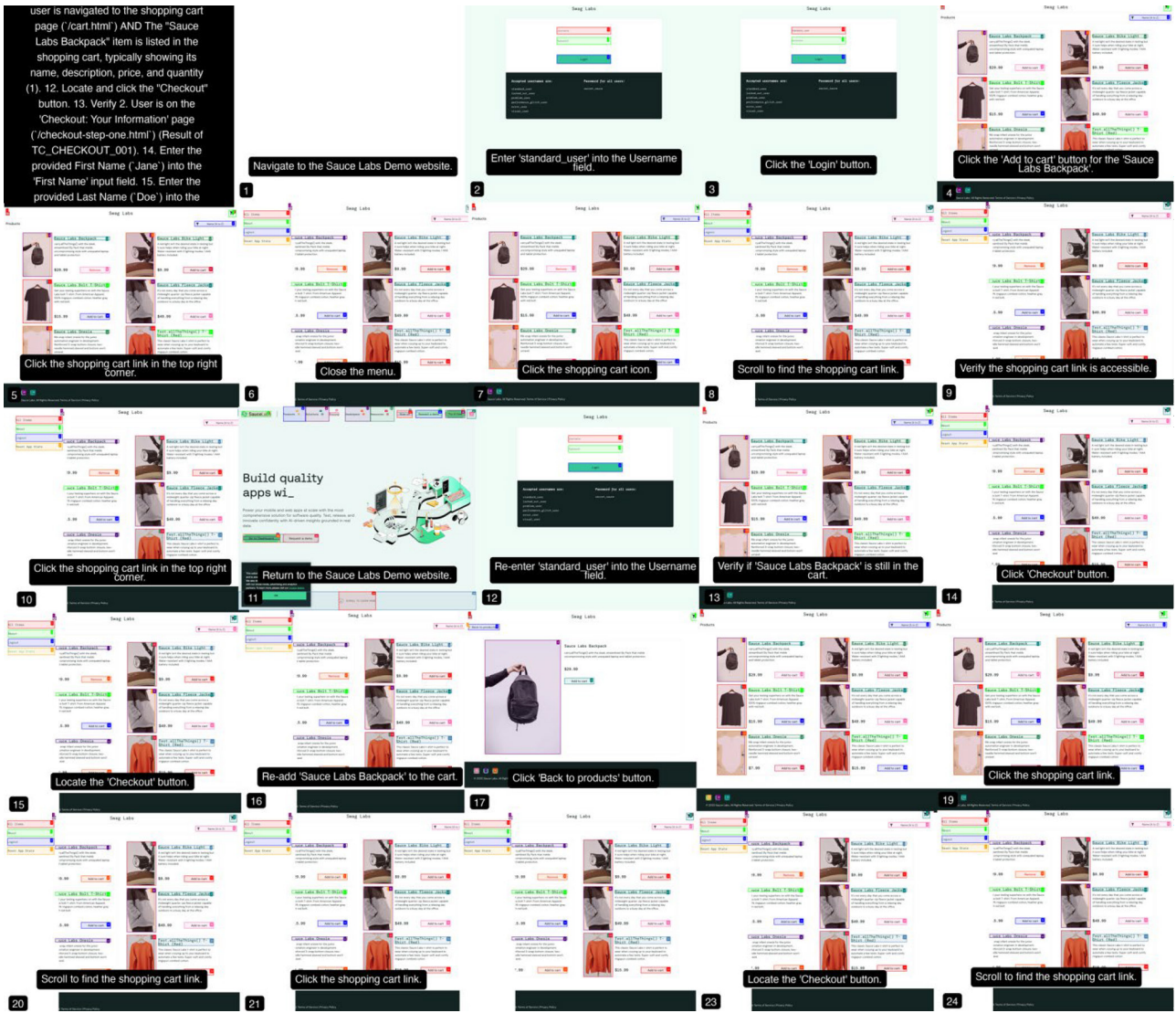

**Fig. 9.** GPT-4o-mini end-to-end flow with reasoning loop

### 3.3. Limitations and future directions of research

Obtained results also revealed limitations for complete adoption of these practices. LLM outputs, especially from smaller models, exhibited inconsistencies in format and quality, often requiring careful prompt engineering or post-processing. The most critical limitation observed was the tendency of AI agents, to deviate from negative test scenarios by attempting to "correct" the flow to achieve a positive outcome, masking potential failures. This underscores a major risk of false negatives and implies that simply checking the final state is insufficient. Furthermore, computational cost and reliance on API availability remain practical constraints.

These limitations guide future research. First, it is necessary the robust, automated methods to evaluate the executability and semantic novelty of AI-generated tests. To empower agentic end-to-end execution, research should also focus on techniques to reliably constrain AI behavior during negative testing as well as reasoning guardrails. Another key area is exploring the effectiveness of domain-specific fine-tuned LLMs for QA tasks compared to general models. Additionally, it is essential to create standardized benchmarks for AI-driven QA tools. These steps will help unlock AI's full potential in enhancing software quality assurance.

### 4. Conclusions

In this paper, the transformative potential of integrating AI-driven tools into modern software quality assurance practices across key verification and validation activities is explored. Transformer-based static analyzers such as CodeBERT narrow the long-standing precision-versus-recall trade-off, boosting F1 scores for critical defect classes by more than 40 percentage points over a SonarQube baseline depending on scenario. Generative models produce fully runnable acceptance tests straight from user stories, achieving 100% criteria coverage while holding cost below 0.005 USD per story in our benchmark. ReAct-style browser agents automate complex end-to-end flows with only 8.3% flaky runs, being on parity with mature scripted suites and indicating a practical path towards more resilient and efficient test automation. This efficiency stems from the AI's ability to interpret requirements, generate test artifacts, and adapt to dynamic UI elements.

Yet practical adoption faces significant challenges including generation of semantically identical coverage, "black box" nature and lack of explainability from state-of-the-art LLMs. Furthermore, the unexpected behavior observed in agentic end-to-end testing. The tendency to correct mutated test cases to match expected results, highlights a significant risk of false negatives emphasizing need of human and machine-based verification of execution results and generation outputs.

Therefore, it is recommended to introduce safeguards. First, it is proposed a safeguard for generated test case quality by combining an LLM-as-a-Judge post-processor with RAG. RAG would source few-shot examples from user stories, documentation or architecture decision records to guide generation and provide context for the judge, aided by embedding-based semantic duplicate pruning. Agent execution should be wrapped in guardrails – token and time caps, mandatory checkpoint assertions, orchestrating a heavy-planner/lightweight-executor model mix to avert reasoning loops and runaway costs. To improve the quality of AI-generated unit tests and enhance system transparency, pairing observation-based unit test generators with a mutation testing feedback loop is recommended. This ensures that only test cases increasing effective coverage are retained. Additionally, all AI components should be required to output a rationale trace, providing essential input for human-in-the-loop review dashboards and addressing explain ability concerns. These controls convert opaque generation into a governed, auditable process.

Beyond these technical challenges and safeguards, another important consideration is the potential deskilling effect on human QA due to utilization of LLMs in explorative test analyses, equivalence partitioning and test scenario design [33]. Lack of firsthand experience in creating test process from scratch may make it impossible to evaluate outputs from LLM when the need arises.

Adoption of each part of discussed framework cuts test-design effort and moves defect discovery further left in the SDLC, chipping away at the estimated 2.4 trillion USD annual cost of poor-quality software. Crucially, it redefines the QA role from script author to curator of knowledge and governor of intelligent systems. Organizations must therefore invest in up-skilling testers as prompt engineers, data stewards, and quality governance owners. With these measures in place, AI-driven tools are positioned not as a replacement for human expertise but as a force multiplier that transforms quality assurance from a late-stage bottleneck into a proactive, economically compelling safety net for continuous delivery.

## Conflict of interest

The authors declare that they have no conflict of interest in relation to this research, including financial, personal, authorship or other, which could affect the research, and its results presented in this article.

## Financing

The research was performed without financial support.

## Data availability

Manuscript has data and source code included as electronic supplementary material. Software implemented for this research as well as all source and generated data is available at GitHub [30].

For Experiment 3.1, user stories are stored in folder *stories*. Test cases generated by LLMs are stored within folder *tc*. Human annotation result is appended at the end of each file within *Verification* section. Code for test-case generation is available in *llm_executor.py*.

Execution logs and screenshots of end-to-end coverage with AI-agents described in Experiment 3.2 is found in *e2elogs* directory. Python script to execute coverage is located in *end_to_end_runner.py*.

## Use of artificial intelligence

The authors have used artificial intelligence technologies within acceptable limits to provide their own verified data, which is described in the research methodology section.

## References

1. Krasner, H. (2022). *The Cost of Poor Software Quality In The Us: A 2022 Report*. Austin. Available at: https://www.it-cisq.org/wp-content/uploads/sites/6/2022/11/CPSQ-Report-Nov-22-2.pdf
2. Cordy, M., Rwemalika, R., Franci, A., Papadakis, M., Harman, M. (2022). FlakiMe: Laboratory-Controlled Test Flakiness Impact Assessment. *Proceedings of the 44th International Conference on Software Engineering.* New York, 982–994. https://doi.org/10.1145/3510003.3510194
3. Gumhold, F. (2022). *How a lack of quality assurance can lead to a loss of $400 million in 37 seconds – ERNI*. Available at: https://www.betterask.erni/how-a-lack-of-quality-assurance-can-lead-to-a-loss-of-400-million-in-37-seconds/
4. *Project Management Institute, A Guide to the Project Management Body of Knowledge and The Standard for Project Management* (2021). Newtown Square: Project Management Institute, Inc.
5. Deshpande, S. A., Deshpande, A. N., Marathe, M. V., Garje, G. V. (2010). Improving Software Quality with Agile Testing. *International Journal of Computer Applications, 1 (22),* 68–73. https://doi.org/10.5120/440-673
6. *Introduction to Software testing* (2025). University of Minnesota Software Engineering Center. Available at: https://www.coursera.org/learn/introduction-software-testing/lecture/ohzH2/introduction
7. *Swag Labs.* Available at: https://www.saucedemo.com/
8. Vokhranov, I., Bulakh, B. (2024). Transformer-based models application for bug detection in source code. *Technology Audit and Production Reserves, 5 (2 (79)),* 6–15. https://doi.org/10.15587/2706-5448.2024.310822
9. *6X improvement over SonarQube – Raising the Maintainability bar.* Available at: https://codescene.com/blog/6x-improvement-over-sonarqube
10. Marjanov, T., Pashchenko, I., Massacci, F. (2022). Machine Learning for Source Code Vulnerability Detection: What Works and What Isn't There Yet. *IEEE Security & Privacy, 20 (5),* 60–76. https://doi.org/10.1109/msec.2022.3176058
11. Roman, A. (2018). *A Study Guide to the ISTQB® Foundation Level 2018 Syllabus.* Springer International Publishing. https://doi.org/10.1007/978-3-319-98740-8
12. Plein, L., Ouédraogo, W. C., Klein, J., Bissyandé, T. F. (2024). Automatic Generation of Test Cases based on Bug Reports: A Feasibility Study with Large Language Models. *Proceedings of the 2024 IEEE/ACM 46th International Conference on Software Engineering: Companion Proceedings,* 360–361. https://doi.org/10.1145/3639478.3643119
13. Ayon, Z. A. H., Husain, G., Bisoi, R., Rahman, W., Osborn, D. T. (2025). *An efficient approach to represent enterprise web application structure using Large Language Model in the service of Intelligent Quality Engineering.* ArXiv. https://doi.org/10.48550/arXiv.2501.06837
14. Li, D., Jiang, B., Huang, L., Beigi, A., Zhao, C., Tan, Z. et al. (2024). *From generation to judgment: Opportunities and challenges of LLM-as-a-judge.* arXiv. https://doi.org/10.48550/arXiv.2411.16594
15. Müller, M., Žunič, G. (2024). *Browser Use: Enable AI to control your browser.* GitHub. Available at: https://github.com/browser-use/browser-use
16. Martin, R. C. (2017). *Clean Architecture: A Craftsman's Guide to Software Structure and Design.* Prentice Hall Press.
17. Daka, E., Fraser, G. (2014). A Survey on Unit Testing Practices and Problems. *2014 IEEE 25th International Symposium on Software Reliability Engineering,* 201–211. https://doi.org/10.1109/issre.2014.11
18. Wang, J., Huang, Y., Chen, C., Liu, Z., Wang, S., Wang, Q. (2024). Software Testing With Large Language Models: Survey, Landscape, and Vision. *IEEE Transactions on Software Engineering, 50 (4),* 911–936. https://doi.org/10.1109/tse.2024.3368208
19. Alshahwan, N., Harman, M., Marginean, A., Tal, R., Wang, E. (2024). Observation-Based Unit Test Generation at Meta. *Companion Proceedings of the 32nd ACM International Conference on the Foundations of Software Engineering,* 173–184. https://doi.org/10.1145/3663529.3663838
20. Alshahwan, N., Chheda, J., Finogenova, A., Gokkaya, B., Harman, M., Harper, I. et al. (2024). Automated Unit Test Improvement using Large Language Models at Meta. *Companion Proceedings of the 32nd ACM International Conference on the Foundations of Software Engineering.* New York, 185–196. https://doi.org/10.1145/3663529.3663839
21. Foster, C., Gulati, A., Harman, M., Harper, I., Mao, K., Ritchey, J. et al. (2025). *Mutation-guided LLM-based test generation at Meta.* arXiv. https://doi.org/10.48550/arXiv.2501.12862
22. Shin, J., Aleithan, R., Hemmati, H., Wang, S. (2024, September). *Retrieval-augmented test generation: How far are we?* arXiv. https://doi.org/10.48550/arXiv.2409.12682
23. Hu, R., Peng, C., Wang, X., Gao, C. (2025). *An LLM-based agent for reliable Docker environment configuration.* arXiv. https://doi.org/10.48550/arXiv.2502.13681
24. Gunawi, H. S., Hao, M., Leesatapornwongsa, T., Patana-anake, T., Do, T., Adityatama, J. et al. (2014). What Bugs Live in the Cloud? A Study of 3000+ Issues in Cloud Systems. *Proceedings of the ACM Symposium on Cloud Computing.* New York. https://doi.org/10.1145/2670979.2670986
25. Vocke, H. (2018). *The Practical Test Pyramid.* Available at: https://martinfowler.com/articles/practical-test-pyramid.html
26. Yao, S., Zhao, J., Yu, D., Du, N., Shafran, I., Narasimhan, K., Cao, Y. (2022). *ReAct: Synergizing reasoning and acting in language models.* arXiv. https://doi.org/10.48550/arXiv.2210.03629

27. Ahlgren, J., Berezin, M., Bojarczuk, K., Dulskyte, E., Dvortsova, I., George, J. et al. (2021). Testing Web Enabled Simulation at Scale Using Metamorphic Testing. *2021 IEEE/ACM 43rd International Conference on Software Engineering: Software Engineering in Practice (ICSE-SEIP),* 140–149. https://doi.org/10.1109/icse-seip52600.2021.00023
28. Ribeiro, M. T., Wu, T., Guestrin, C., Singh, S. (2020). Beyond Accuracy: Behavioral Testing of NLP Models with CheckList. *Proceedings of the 58th Annual Meeting of the Association for Computational Linguistics.* Stroudsburg, 4902–4912. https://doi.org/10.18653/v1/2020.acl-main.442
29. John-Mathews, J.-M. (2022). *How to test Machine Learning Models?* Metamorphic testing. Available at: https://www.giskard.ai/knowledge/how-to-test-ml-models-4-metamorphic-testing
30. *Data for paper AI-Driven Testing Tools in Modern Quality Assurance: Benefits, Challenges, and Future Directions.* Available at: https://github.com/igor-pysmennyi-kpi/qa-ai-overview-paper-2025
31. Micco, J. (2016). *Flaky Tests at Google and How We Mitigate Them.* Google Testing Blog. Available at: https://testing.googleblog.com/2016/05/flaky-tests-at-google-and-how-we.html
32. Valmeekam, K., Olmo, A., Sreedharan, S., Kambhampati, S. (2022). Large Language Models Still Can't Plan (A Benchmark for LLMs on Planning and Reasoning about Change). *iNeurIPS 2022 Foundation Models for Decision Making Workshop, New Orleans.* Available at: https://openreview.net/forum?id=wUU-7XTL5XO
33. Lee, H.-P. (Hank), Sarkar, A., Tankelevitch, L., Drosos, I., Rintel, S., Banks, R., Wilson, N. (2025). The Impact of Generative AI on Critical Thinking: Self-Reported Reductions in Cognitive Effort and Confidence Effects from a Survey of Knowledge Workers. *Proceedings of the 2025 CHI Conference on Human Factors in Computing Systems,* 1–22. https://doi.org/10.1145/3706598.3713778

✉*Ihor Pysmennyi, PhD, Senior Lecturer, Department of System Design, National Technical University of Ukraine "Igor Sikorsky Kyiv Polytechnic Institute", Kyiv, Ukraine, e-mail: Ihor.pismennyy@gmail.com, ORCID: https://orcid.org/0000-0001-7648-2593*

------------------------

*Roman Kyslyi, PhD, Senior Lecturer, Department of System Design, National Technical University of Ukraine "Igor Sikorsky Kyiv Polytechnic Institute", Kyiv, Ukraine, ORCID: https://orcid.org/0000-0002-8290-9917*

------------------------

*Kyrylo Kleshch, PhD, Assistant, Department of System Design, National Technical University of Ukraine "Igor Sikorsky Kyiv Polytechnic Institute", Kyiv, Ukraine, ORCID: https://orcid.org/0009-0006-8133-3086*

------------------------

✉*Corresponding author*